\documentclass[conference]{IEEEtran}
\usepackage{cite}
\usepackage{amsmath,amssymb,amsfonts}
\usepackage{algorithmic}
\usepackage{graphicx}
\usepackage{textcomp}
\usepackage[table]{xcolor}
\usepackage{multicol,multirow,pifont,amssymb,subcaption}

\newcommand{\xmark}{\ding{55}}%

\def\BibTeX{{\rm B\kern-.05em{\sc i\kern-.025em b}\kern-.08em
    T\kern-.1667em\lower.7ex\hbox{E}\kern-.125emX}}

\newcommand{\copyrightnotice}{
\begin{minipage}{\textwidth}
    \centering
    {\small © 2025 IEEE.  Personal use of this material is permitted.  Permission from IEEE must be obtained for all other uses, in any current or future media, including reprinting/republishing this material for advertising or promotional purposes, creating new collective works, for resale or redistribution to servers or lists, or reuse of any copyrighted component of this work in other works.}
\end{minipage}
}

\begin{document}

\thispagestyle{empty} % Ensure no headers/footers appear
\copyrightnotice
\newpage % Force everything else to the next page
\setcounter{page}{1} % Reset page numbering (first content page starts at 1)

\title{Efficient Finetuning for Dimensional Speech Emotion Recognition in the Age of Transformers}

% Efficient Finetuning Strategies for Wav2Vec 2.0 in Dimensional Speech Emotion Recognition: A Comparative Study}
% 1\textsupserscript{st}
\author{\IEEEauthorblockN{Aneesha Sampath\IEEEauthorrefmark{1}}
\IEEEauthorblockA{\textit{Computer Science and Engineering} \\
\textit{University of Michigan}\\
Ann Arbor, USA \\
saneesha@umich.edu}
\and
\and
\IEEEauthorblockN{James Tavernor\IEEEauthorrefmark{1}}
\IEEEauthorblockA{\textit{Computer Science and Engineering} \\
\textit{University of Michigan}\\
Ann Arbor, USA \\
tavernor@umich.edu}
\and
% 3\textsuperscript{rd} 
\IEEEauthorblockN{Emily Mower Provost}
\IEEEauthorblockA{\textit{Computer Science and Engineering} \\
\textit{University of Michigan}\\
Ann Arbor, USA \\
emilykmp@umich.edu}

\thanks{\IEEEauthorrefmark{1}Equal contribution.}
}
% \and
% \and
% \and

\maketitle

\begin{abstract}
Accurate speech emotion recognition is essential for developing human-facing systems. Recent advancements have included finetuning large, pretrained transformer models like Wav2Vec 2.0.  However, the finetuning process requires substantial computational resources, including high-memory GPUs and significant processing time. As the demand for accurate emotion recognition continues to grow, efficient finetuning approaches are needed to reduce the computational burden. Our study focuses on \textit{dimensional} emotion recognition, predicting attributes such as activation (calm to excited) and valence (negative to positive). We present various finetuning techniques, including full finetuning, partial finetuning of transformer layers, finetuning with mixed precision, partial finetuning with caching, and low-rank adaptation (LoRA) on the Wav2Vec 2.0 base model. We find that partial finetuning with mixed precision achieves performance comparable to full finetuning while increasing training speed by 67\%. Caching intermediate representations further boosts efficiency, yielding an 88\% speedup and a 71\% reduction in learnable parameters. We recommend finetuning the final three transformer layers in mixed precision to balance performance and training efficiency, and adding intermediate representation caching for optimal speed with minimal performance trade-offs. These findings lower the barriers to finetuning speech emotion recognition systems, making accurate emotion recognition more accessible to a broader range of researchers and practitioners.

% Our results show that combining partial finetuning with mixed precision training results in performances that are not significantly different from  full finetuning, while also increasing training speed by 67\%. We further improve training efficiency by caching intermediate frozen Wav2Vec 2.0 layers during mixed precision partial finetuning. This results in an 88\% speedup and a 71\% reduction in learnable parameters. 
\end{abstract}


\section{Introduction}
Speech emotion recognition plays a critical role in enabling systems to detect the emotional state of users. Recent advancements in speech emotion recognition have centered on the finetuning of pretrained transformer models \cite{wagner2023dawn} such as Wav2Vec 2.0 \cite{baevski2020wav2vec}. However, finetuning large models requires substantial computational resources, including significant memory and processing time, which are not easily available to many researchers and practitioners. As the demand for emotion recognition grows, there is an increasing need for more efficient finetuning approaches to reduce the computational burden.

Most prior work in finetuning Wav2Vec 2.0 for \textit{dimensional} speech emotion recognition, which focuses on predicting continuous emotion attributes such as activation (ranging from calm to active) and valence (ranging from negative to positive) \cite{harmon2017importance, russell1979affective, lotfian2017buildingpodcast}, sets the training batch size to at least 32 \cite{wagner2023dawn, triantafyllopoulos2022probing}. The combination of this batch size and the large parameter count of Wav2Vec 2.0 demands substantial GPU memory. Yet, a large batch size is essential because the standard loss function for dimensional emotion recognition is batch-dependent. Common GPUs such as the NVIDIA GTX 1080 Ti and RTX 2080 Ti yield ``out-of-memory" errors when finetuning Wav2Vec 2.0 base with the standard training configurations. Therefore, finetuning even the base model is not possible without significant computational resources, limiting its use to practitioners with high-end GPUs. This highlights the critical need for exploration of efficient finetuning techniques that can maintain performance while alleviating the computational demands. 

% In this work, we present a novel and comprehensive comparison of finetuning methods for Wav2Vec 2.0, which, to the best of our knowledge, is the first work to systematically explore and compare partial finetuning of transformer layers, mixed precision training \cite{micikevicius2017mixed}, caching intermediate representations, and LoRA (Low-Rank Adaptation) \cite{hu2021lora} for dimensional speech emotion recognition. 

In this work, we present a comprehensive comparison of finetuning methods for Wav2Vec 2.0. To the best of our knowledge, this is the first work to systematically evaluate and compare partial finetuning of transformer layers, mixed precision training \cite{micikevicius2017mixed}, caching intermediate representations, and LoRA (Low-Rank Adaptation) \cite{hu2021lora} within a single study for dimensional speech emotion recognition, offering guidelines on resource requirements to achieve state-of-the-art results. We find that partial finetuning of the Wav2Vec 2.0 base model achieves comparable performance to full finetuning, showing no statistically significant differences. Furthermore, we find that combining partial finetuning with mixed-precision training results in no significant difference compared to full finetuning with a 67\% speedup. We combine this approach with caching to further speedup training by 88\% over full finetuning. Notably, the partial finetuning approach can be executed on GPUs with lower memory, making it accessible to a wider range of researchers and practitioners. 

\section{Related Work}
\label{sec:related}

\subsection{Dimensional Emotion Recognition}
We focus on finetuning Wav2Vec 2.0 for dimensional emotion recognition. The \textit{dimensional} emotion theory posits that emotion can be described by core attributes like valence (negative to positive) and activation (calm to active) \cite{harmon2017importance, russell1979affective}. 
%This theory allows for the joint modeling of valence and activation, allowing for the representation of nuanced emotions. 

\subsection{Wav2Vec 2.0}

Wav2Vec 2.0 is a transformer-based model that learns contextualized representations from unlabeled raw audio data through self-supervised learning \cite{baevski2020wav2vec}. It consists of a CNN-based feature encoder, transformer-based context network, and quantization module to discretize the feature encoder output, with a convolution layer before the transformer layers to learn positional embeddings \cite{baevski2020wav2vec}. Wav2Vec 2.0 has been finetuned for tasks beyond automatic speech recognition, including speaker recognition \cite{vaessen2022fine, yi2020applying}, speaker verification \cite{chen2022large, wang2021fine}, and speech emotion recognition \cite{pepino2021emotion, wang2021fine, chen2023exploring, wagner2023dawn}.

Wav2Vec 2.0 embeddings have been shown to outperform traditional, non-deep-learning-based speech-emotion features \cite{pepino2021emotion}, such as eGeMAPS \cite{eyben2015geneva} and spectograms. As a result, deep embeddings have become the foundation for many modern speech emotion recognition systems \cite{pepino2021emotion, wang2021fine, chen2023exploring, wagner2023dawn}.

\subsection{Efficient Transformer Model Finetuning}

% The pretraining-finetuning paradigm, popularized by BERT \cite{devlin2018bert}, allows a model to leverage general knowledge, such as language and syntax, though a task-agnostic objective. Finetuning involves only the \textit{adjustment} of model parameters to specialize the model for a task. However, finetuning models like BERT base (110M parameters) and Wav2Vec 2.0 base (95M parameters) remains computationally expensive and often infeasible on common hardware. Moreover, using Wav2Vec 2.0 with raw audio input demands significant GPU memory. Recent work has further analyzed the effectiveness of training all layers in both BERT and Wav2Vec 2.0.

Pretraining allows models to learn general knowledge, such as language and syntax, through a task-agnostic objective. Finetuning \textit{adjusts} pretrained model parameters for specific tasks. However, finetuning Wav2Vec 2.0 base (95M parameters) is computationally expensive and often impractical on common GPUs due to the high memory demands of raw audio input. Understanding the roles of individual layers can provide a foundation for exploring efficient finetuning strategies.

% Recent studies have examined the role of individual layers in Wav2Vec 2.0.

% In natural language processing, Tenney et al. found that the initial transformer layers of BERT primarily encode syntax, while the final layers capture semantics \cite{tenney2019bert}, indicating that finetuning the final layers may suffice for state-of-the-art performance. Furthermore, Lee et al. found that finetuning just one fourth of BERT's transformer layers can achieve 90\% of the performance of full finetuning across various natural language processing tasks, with partial finetuning outperforming full finetuning for sentiment classification \cite{lee2019would}.

Pasad et al. performed a layer-wise analysis of Wav2Vec 2.0, showing that the early transformer layers encode acoustics, middle layers capture phonetics, and upper layers focus on semantics \cite{pasad2021layer}. When finetuning Wav2Vec 2.0 base for automatic speech recognition, they found that the early layers remained highly correlated with the pretrained checkpoint, while the final 3 to 4 layers encoded task-specific information. This supports an analysis into partial, rather than full, finetuning.

% Our work  focuses on the effectiveness of finetuning only the final transformer layers compared to full finetuning.

% Efficient finetuning has been less studied in speech processing models. 
Prior work has explored modifications to the Wav2Vec 2.0 architecture to improve speed and performance in speech recognition \cite{wu2022performance}. Additionally, prior work has explored the partial finetuning of Wav2Vec 2.0 by freezing the CNN feature encoder and finetuning only the transformer layers \cite{wang2021fine}, as recommended by the Wav2Vec 2.0 authors \cite{baevski2020wav2vec}. Wagner et al. found that freezing the transformer layers and training only the output heads resulted in a substantial decrease in emotion performance, indicating that finetuning the transformer layers is necessary for achieving state-of-the-art results \cite{wagner2023dawn}. To the best of our knowledge, this is the first work to examine the effect of partial finetuning within the transformer layers and, more generally, efficient training techniques, for dimensional speech emotion recognition. For the remainder of the paper, we freeze the CNN feature encoder and define ``partial finetuning" as the selective freezing of transformer layers.

Mixed precision training \cite{micikevicius2017mixed} has become a common approach for efficiently finetuning large models. It involves storing weights, gradients, and activations in half-precision, but maintaining single-precision copies of weights to \textit{accumulate} gradients. Previous work has combined mixed precision with other techniques to reduce GPU memory requirements for Wav2Vec 2.0 training in automatic speech recognition \cite{lugo2024sustainable}. To the best of our knowledge, this is the first work to apply mixed precision training to dimensional speech emotion recognition.

We also explore the Parameter-Efficient Finetuning technique LoRA (low-rank approximation), which freezes all pretrained model weights and adds small rank decompositional matrices to model layers (typically attention layers) during training \cite{hu2021lora}. Feng et al. found that LoRA finetuning was beneficial for categorical speech emotion classification \cite{feng2023peft}. To the best of our knowledge, this is the first work to explore LoRA finetuning for dimensional speech emotion recognition.

\subsection{Gradient Checkpointing}
Gradient checkpointing reduces memory consumption by storing only a subset of model outputs at designated checkpoints, rather than all intermediate outputs. These are recomputed during the backward pass \cite{8855345}. While checkpointing does not affect training accuracy, it increases training time by about 30\% compared to full finetuning \cite{sohoni2019low}. Therefore, we do not compare gradient checkpointing, and instead investigate faster methods, such as caching intermediate model outputs from frozen layers during partial finetuning (Section~\ref{sec:caching}).

% Gradient checkpointing decreases memory usage by increasing computation time. It only outputs at designated checkpoints that are saved during the model's forward pass, discarding values in between checkpoints.  During the backward pass, these are recomputed \cite{8855345}. We do not compare gradient checkpointing as it should not significantly change model outputs and is slower than full finetuning.  Instead, we investigate faster methods, like caching intermediate model outputs from frozen layers when performing partial finetuning (Section~\ref{sec:caching}).

\section{Dataset}
\label{sec:dataset}

The MSP-Podcast dataset contains emotional, non-acted speech from podcasts, annotated for categorical emotions and dimensions of activation, valence, and dominance \cite{lotfian2017buildingpodcast}. We focus on predicting activation and valence, using release 1.11 (May 31, 2023), which includes 151,654 segments from 2,172 speakers across 237 hours of audio. We train on the `Train' split and evaluate on the `Test1' split, and select the model with the best performance on the `Development' split. We scale the labels from 1 to 7 to 0 to 1, as in \cite{wagner2023dawn}. We use the original lengths of the speech samples for all predictions.

\section{Method}
\label{sec:method}

\subsection{Architecture}
In all approaches, we use the standard Wav2Vec 2.0 architecture. We freeze the CNN feature extractor, as suggested by the Wav2Vec 2.0 authors \cite{baevski2020wav2vec}, and use Huggingface checkpoint \textit{facebook/wav2vec2-base}\footnote{https://huggingface.co/facebook/wav2vec2-base}. We apply mean pooling over the hidden states of the final transformer layer \cite{wagner2023dawn}, apply dropout ($p=0.2$), and pass the output into a multitask output consisting of two task-specific heads (activation and valence).

\subsection{Finetuning Approaches}

\begin{figure}[t]
    \centering
    \begin{subfigure}[t]{0.12\textwidth}
        \includegraphics[width=\textwidth]{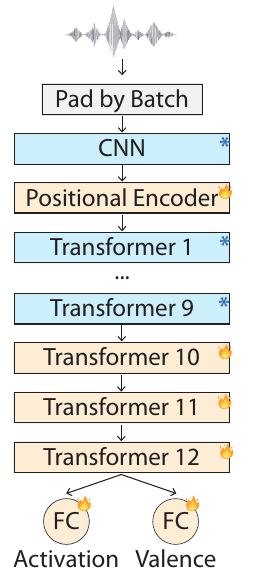}
        \caption{Non-caching}
        \label{fig:no_cache}
    \end{subfigure}
    % \hfill
    \begin{subfigure}[t]{0.23\textwidth}
        \includegraphics[width=\textwidth]{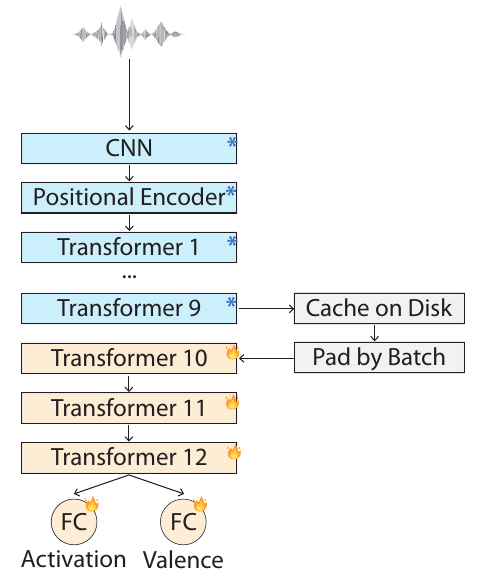}
        \caption{Caching}
        \label{fig:cache}
    \end{subfigure}
    \caption{Non-caching vs caching approach for partial finetuning (three-layer). The blue layers with the snowflake are frozen layers, whereas the orange layers with the fire are trainable layers. `FC' represents fully connected layers.}
    \label{fig:diagram}
\end{figure}

% We experiment with the following finetuning approaches. 
In the \textbf{full finetune}, we finetune all 12 transformer layers. In the \textbf{mixed precision full finetune}, we finetune all 12 transformer layers with mixed precision. In the \textbf{partial finetune}, we freeze the first $12-n$ and finetune the final $n$ transformer layers, where $1 \leq n \leq 3$ (Figure \ref{fig:diagram}a). In the \textbf{mixed precision partial finetune}, we finetune the final $n$ transformer layers with mixed precision. In the \textbf{LoRA finetune}, we freeze all pretrained model weights and apply LoRA to the query and value projections in all 12 transformer layers, with LoRA paramaters $\text{rank}=8$, $\text{alpha}=16$, $\text{dropout}=0.1$.

% \begin{figure}[t]
%     \centering
%     \begin{minipage}{0.43\columnwidth} % Adjust width as needed
%         \centering
%         \includegraphics[width=0.65\columnwidth]{architecture.pdf}
%         \caption*{(a) Non-caching} % Use \caption* to avoid extra spacing issues
%         \label{fig:no_cache}
%     \end{minipage}
%     \begin{minipage}{0.52\columnwidth}
%         \centering
%         \includegraphics[width=\columnwidth]{architecture_caching.pdf}
%         \caption*{(b) Caching}
%         \label{fig:cache}
%     \end{minipage}
%     \caption{Non-caching vs caching approach for partial finetuning (three-layer). The blue layers with the snowflake are frozen layers, whereas the orange layers with the fire are trainable layers. `FC' represents fully connected layers.}
%     \label{fig:diagram}
% \end{figure}

In these experimental setups, audio samples are processed in batches. Thus, we must zero-pad the audio, which adds silence to the end of the raw waveforms so that all samples in a batch are equal in length. For non-caching experiments, we follow the standard protocol of zero-padding audio \textit{before} it enters the model, shown in Figure \ref{fig:diagram}a.  We discuss zero-padding for caching experiments in Section~\ref{sec:caching}.
%% Explain zero padding and why it is needed

\subsection{Caching}
\label{sec:caching}
Due to the large number of parameters in Wav2Vec 2.0, the computation time is significant even when most layers are frozen. However, it is important to note that the output of frozen layers does not change for a given input audio sample. Therefore, we propose storing the model output from the frozen layers for a one-time cost. In this case, we exchange increased disk usage for decreased computation time and memory usage. Additionally, we must freeze the positional embedding layer in the caching approach since it occurs before transformer layers (Figure \ref{fig:diagram}b).

%% Relate back to what you are talking about earlier

% treat each audio sample as a `batch'

This involves three steps: 1) process each audio sample individually (batch size = 1), creating and caching representations of that sample up to the final frozen layer (e.g., layer 9 in Figure~\ref{fig:diagram}b), 2) when preparing to train the model, create batches by loading  subsets of the cached representations, 3) train the system as in the non-caching setup.  Like in the non-caching case, samples in a batch must be of the same length.  But, in this case, we do not know the length in advance.  Therefore, we create a special \textbf{caching padding strategy}. Instead of zero-padding an audio sample before it enters the model, we must zero-pad its cached representation.  However, Wav2Vec 2.0 base processes zero-padding without an attention mask, meaning the model does not differentiate between padding and silence. This is not a problem when the data are processed from raw audio, but may introduce performance changes when the \textit{representations} are zero-padded in the middle of a set of transformer layers.  We investigate the effects of this on downstream performance by presenting both the efficiency of caching and the performance of the resulting systems.

% Wav2Vec 2.0 base processes zero-padding without an attention mask, treating padded and non-padded inputs the same. In the non-caching approach, inputs are zero-padded before the model processes the input (Figure \ref{fig:diagram}a), but with caching, it is introduced after the $n$th transformer layer (Figure \ref{fig:diagram}b). Additionally, the cached representations are fixed with the model in either \textit{eval} or \textit{train} mode, which may impact model outputs. 

% %% Add some context to this
% We fix cached representations in \textit{train} mode, and freeze the positional embedding layer since it occurs before the transformer layers. [remove if no difference in result]

\subsection{Model Training}

We compute the loss separately for valence and activation and average them.  We train the model for five epochs, and select the model with the best performance on the development set.  We use the AdamW optimizer \cite{loshchilov2017decoupled} with Concordance Correlation Coefficient (CCC) loss \cite{lawrence1989concordance}, the standard loss for dimensional emotion recognition \cite{wagner2023dawn, parthasarathy2017jointly, triantafyllopoulos2022probing}, and  a fixed learning rate of $1\mathrm{e}{-4}$ with batch size 32, as in \cite{wagner2023dawn}. 

%\subsection{Experimental Setup}

%We run each training configuration across 5 random seeds and report the mean performance and standard deviation, and [UPDATE THIS] test statistical significance using a t-test, asserting significance when $p < 0.05$.

\subsection{System Configuration}

We run caching experiments on a single NVIDIA RTX 2080 Ti and all other experiments on a high-performance cluster with a single NVIDIA A40. We cannot perform caching experiments on the high-performance cluster on the A40 since the caching approach requires significant disk space (113GB per cached layer). We test for out-of-memory errors on the NVIDIA GTX 1080 Ti, in addition to the RTX and A40.  We train for one epoch on an GTX 1080 Ti (11GB), RTX 2080 Ti (11GB), and A40 (48GB), and report if the finetuning approaches result in out-of-memory errors.

%\subsection{Out-of-Memory Check}

%We train for one epoch on an NVIDIA GTX 1080 Ti (11GB), RTX 2080 Ti (11GB), and A40 (48GB), and report if the finetuning approaches result in out-of-memory errors.

\section{Results}
\label{sec:results}

We run each training configuration across five random seeds.  We report the mean performance and standard deviation, and test statistical significance. We assert significance when $p < 0.05$. We first compare the results over the different finetuning methods, focusing only on non-caching experiments. We use a one-way ANOVA to determine if performance differences across finetuning approaches are significant. If so, we follow this analysis with t-tests across the non-caching experiments, using the Bonferroni correction. Next, we compare non-caching to caching experiments to assess whether caching significantly hampers emotion recognition performance. Since this involves a single comparison (e.g., partial finetuning with or without caching), the Bonferroni correction is not applied.

We discuss all approaches across both activation and valence. We report CCC and the associated training time in hours in Table \ref{tab:combined_results_table}. The results for standard finetuning, 12-layer finetune in single-precision (the baseline), are shaded.

 % asserting significance when $p < 0.05$.

% \input{tables/shared_table}
% \input{tables/two_tables}
\begin{table*}[h]
%\caption{Wav2Vec 2.0 results and training time. We report the number of transformer layers trained (`Layers'), activation and valence CCC on MSP-Podcast ``Test1" split (`Act' and `Val'), wall-clock training time in hours for 5 epochs (`Time'), and number of trainable parameters ('Params'). We report results for single precision (SP) and mixed precision (MP) separately. * indicates statistically significant improvement, and $\dagger$ indicates statistically significant decrease from full 12-layer finetune. Significance for caching results are reported relative to their partial finetuning counterpart. We report `NS' when the difference between non-caching and its caching counterpart is nonsignificant. These experiments were conducted on NVIDIA A40 (non-caching) and RTX 2080 Ti (caching) GPUs.}
\caption{Results and training time for the number of transformer layers trained (`Layers'), activation (`Act') and valence (`Val') CCC, wall-clock training time in hours for 5 epochs (`Time'), and trainable parameters (`Params'), for single precision (SP) and mixed precision (MP). Non-caching results are compared to full finetuning SP (shaded). Caching results are compared to the non-caching equivalent. $\dagger$ indicates significant decrease. `NS' (not significant) shows that caching had no significant impact on performance. Experiments were conducted on NVIDIA A40 (non-caching) and RTX 2080 Ti (caching) GPUs.}
\label{tab:combined_results_table}
\centering
\begin{tabular}{l c c c c c||c c c c}
\hline
\multicolumn{6}{c||}{\textbf{Non-Caching Results}} & \multicolumn{4}{c}{\textbf{Caching Results}} \\

\textbf{Layers} & \textbf{Precision} & \textbf{Act} & \textbf{Val} & \textbf{Time (h)} & \textbf{Params} & \textbf{Act} & \textbf{Val} & \textbf{Time (h)} & \textbf{Params} \\

\hline
\multirow{1}{*}{LoRA} & SP & 0.623±0.02 & 0.506±0.01$\dagger$ & 2.476±0.02 & 300K & -- & -- & -- & -- \\
\hline
\multirow{2}{*}{1} & SP & 0.622±0.02 & 0.458±0.01$\dagger$ & 2.443±0.002 & \multirow{2}{*}{12M} & 
NS % 0.631±0.01 
& NS % 0.461±0.01 
& 0.353±0.002 & \multirow{2}{*}{7M}\\
& MP & 0.625±0.01 & 0.462±0.01$\dagger$ & 0.965±0.01 && 
NS % 0.631±0.01 
& NS % 0.461±0.01 
& 0.155±0.01 & \\
\hline

\multirow{2}{*}{2} & SP & 0.638±0.01 & 0.508±0.01$\dagger$ & 2.477±0.01 & \multirow{2}{*}{19M}& 
NS % 0.637±0.02 
& NS % 0.507±0.004 
& 0.689±0.002 & \multirow{2}{*}{14M} \\

& MP & 0.645±0.01 & 0.507±0.004$\dagger$ & 0.982±0.003 && 
NS % 0.634±0.01
& NS % 0.508±0.01 
& 0.253±0.002 & \\
\hline

\multirow{2}{*}{\textbf{3}} & SP & 0.648±0.01 & 0.565±0.002 & 2.519±0.02 & \multirow{2}{*}{26M} & 
NS % 0.644±0.01 
& 0.558±0.003$\dagger$ & 1.023±0.002 & \multirow{2}{*}{21M}\\
& \textbf{MP} & \textbf{0.655±0.01} & \textbf{0.568±0.01} & 0.991±0.003 && 0.639±0.01$\dagger$ & 
NS % 0.555±0.01 
& 0.353±0.002 & 
\\
\hline

\multirow{2}{*}{12} & \cellcolor{gray!20}SP & \cellcolor{gray!20}0.637±0.01 & \cellcolor{gray!20}0.567±0.02 & \cellcolor{gray!20}2.980±0.01 & \multirow{2}{*}{90M} & \multirow{2}{*}{--} & \multirow{2}{*}{--} & \multirow{2}{*}{--} & \multirow{2}{*}{--}\\
 & MP & 0.641±0.01 & 0.566±0.01 & 1.145±0.01 & & \\
%& & & & & \\
\hline
\end{tabular}
\end{table*}

%% Bonferroni correction -- pvalues
% full_act_fp16: reject: False, corrected pval: 0.9970283310062391
% full_val_fp16: reject: False, corrected pval: 0.9984327302286997
% froz9_act_fp32: reject: False, corrected pval: 0.7901513067898108
% froz9_val_fp32: reject: False, corrected pval: 0.9984327302286997
% froz9_act_fp16: reject: False, corrected pval: 0.1499294651425318
% froz9_val_fp16: reject: False, corrected pval: 0.9984327302286997
% froz10_act_fp32: reject: False, corrected pval: 0.9984327302286997
% froz10_val_fp32: reject: True, corrected pval: 0.002730301752521819
% froz10_act_fp16: reject: False, corrected pval: 0.8152944518400996
% froz10_val_fp16: reject: True, corrected pval: 0.00090209588702665
% froz11_act_fp32: reject: False, corrected pval: 0.7939124529829344
% froz11_val_fp32: reject: True, corrected pval: 2.26032168945237e-05
% froz11_act_fp16: reject: False, corrected pval: 0.7939124529829344
% froz11_val_fp16: reject: True, corrected pval: 3.554344296527137e-05
% lora_act: reject: False, corrected pval: 0.7939124529829344
% lora_val: reject: True, corrected pval: 0.0018728458532336536

\subsection{Full and Partial Finetune: Single Precision}
Our experiments show that finetuning the final three transformer layers is as effective as full finetuning, as seen in Table \ref{tab:combined_results_table} in the `SP' rows. There is no significant difference between the two approaches for either activation or valence. This finding aligns with prior work, which showed that the final layers of Wav2Vec 2.0 base encode task-specific information, while the initial layers remain highly correlated with the pretrained checkpoint \cite{pasad2021layer}. Importantly, the three-layer approach is more efficient, as it requires training only about 28\% of the Wav2Vec 2.0 base parameters, compared to 96\% in full finetuning.  

We find that finetuning only one or two layers results in a significant decrease in valence performance compared to full finetuning. However, the difference in activation performance is not significant. This suggests that finetuning one or two transformer layers is sufficient for activation, but not  valence.

%We hypothesize that full finetuning may unnecessarily alter parameters related to general speech, while retaining 9 layers of pretraining knowledge and tuning only 3 for emotion recognition is sufficient. 

% Additionally, activation performance remains consistent across all setups, indicating that more efficiency techniques, such as lower LoRA rank or tuning fewer transformer layers, could be considered when optimizing for activation.

\subsection{Full and Partial Finetune: Mixed Precision}

The results in the previous section demonstrated that partial finetuning is as effective as full finetuning. In this section, we speed up training with mixed precision and investigate the how the performance of the system changes. The results are in the `MP' rows in Table \ref{tab:combined_results_table}.
%The 3-layer mixed precision finetune has no signficant difference in activation and valence performance when compared to the full finetune. It also results in an average 67\% speedup, making it an optimal training technique. 
We find that mixed precision partial finetuning of the final three layers does not significantly affect either activation or valence performance, compared to full finetuning, and offers a 67\% speedup. As with single-precision, finetuning only one or two transformer layers in mixed precision is sufficient for activation, but not for valence.

% here?  Then partial, then caching...
\subsection{LoRA Finetune}
LoRA finetuning has the fewest trainable parameters (300K vs. 90M in full finetuning). Activation performance shows no significant difference between full and LoRA finetuning, despite the large reduction in trainable parameters. Valence performance is significantly worse (0.064 CCC decrease). %Future work could explore increasing the LoRA rank to increase trainable parameters, or the combination of partial finetuning with LoRA by finetuning the final three transformer layers for valence while applying LoRA for activation.

\subsection{Proposed Partial Finetune: Caching with Single Precision and Mixed Precision }

The caching approach showed the largest speedup compared to the full finetune (`Caching Results' in Table \ref{tab:combined_results_table}). Mixed precision caching approaches also offer a greater than 64\% speedup compared to relative non-caching approaches. We find that in one and two layer partial finetuning with caching, the performance of both activation and valence is not significantly different from non-caching in both single and mixed precision (Table~\ref{tab:combined_results_table}, NS indicates not significant). However, it is significantly worse than non-caching in the three-layer partial finetune for activation in mixed precision (0.016 CCC decrease), and for valence in single precision (0.007 CCC decrease). This is likely due to the caching padding strategy, which has propagating effects as it is applied to subsequent trainable layers. The difference in performance could also result from using different GPUs (due to the different system configuration needs, A40 for non-caching and RTX 2080 Ti for caching). However, the relatively similar performance of the models suggests efficiency can be added to the training process, with only minor changes in performance.
%Caching performance in both activation and valence is not significantly different from non-caching in the 1 and 2 layer partial finetune. However, it is significantly worse than non-caching in the 3-layer partial finetune for activation in mixed precision, and in valence for single precision. This may be due to the effect of the caching padding strategy change, which may propagate as we introduce more trainable layers. It could also result from using different GPUs (A40 for non-caching and RTX 2080 Ti for caching). Still, caching remains comparable to full finetuning while offering substantial speedup.

\subsection{Out-of-Memory}

All proposed training approaches, except full finetuning, can be executed on lower-memory GPUs without causing out-of-memory errors, as shown in Table \ref{tab:oom_check}. Specifically, the three-layer mixed precision partial finetuning approach is feasible on 11GB memory GPUs, making state-of-the-art performance more accessible, even with limited computational resources.

\begin{table}[t]
    \caption{GPUs and out-of-memory occurrences. `MP' indicates mixed precision training. `\xmark' indicates out-of-memory.} 
    \label{tab:oom_check}
    \centering
    \begin{tabular}{l c c c}
        \hline
         & RTX 2080 Ti & GTX 1080 Ti & A40 \\
        \hline
        Full            & \xmark & \xmark & \checkmark \\
        Full MP & \xmark & \xmark & \checkmark \\
        Partial         & \checkmark & \checkmark & \checkmark \\
        Partial MP      & \checkmark & \checkmark & \checkmark \\
        LoRA            & \checkmark & \checkmark & \checkmark \\
        \hline
    \end{tabular}
\end{table}

\section{Conclusion}
\label{sec:conclusion}

In this paper, we explore efficent methods for finetuning Wav2Vec 2.0 base for dimensional speech emotion recognition, including full finetuning, partial finetuning of transformer layers, mixed precision training, LoRA finetuning, and caching intermediate representations. Partial finetuning of the final three transformer layers performs comparably to full finetuning. When combined with mixed precision, it offers similar performance, with a 67\% speedup, and can be executed on lower-memory GPUs. Caching intermediate representations further accelerates mixed precision partial finetuning by 88\% compared to full finetuning. Notably, all setups except full finetuning can run on lower-memory GPUs (e.g., RTX 2080 Ti). Our results suggest that partial finetuning, combined with strategies like mixed precision training and caching, is effective for achieving state-of-the-art performance while being fast and resource-efficient. 
%In all setups, the models tended to overfit faster for activation than valence, suggesting that valence requires more parameters or training epochs to reach optimal performance. 
Future work will include finetuning focused on the differences between activation and valence performance, and can validate our findings in other popular pretrained models commonly used for speech emotion recognition, such as WavLM \cite{chen2022wavlm} and HuBERT \cite{hsu2021hubert}.

\bibliographystyle{IEEEtran}
\bibliography{strings}

\begin{thebibliography}{10}
\providecommand{\url}[1]{#1}
\csname url@samestyle\endcsname
\providecommand{\newblock}{\relax}
\providecommand{\bibinfo}[2]{#2}
\providecommand{\BIBentrySTDinterwordspacing}{\spaceskip=0pt\relax}
\providecommand{\BIBentryALTinterwordstretchfactor}{4}
\providecommand{\BIBentryALTinterwordspacing}{\spaceskip=\fontdimen2\font plus
\BIBentryALTinterwordstretchfactor\fontdimen3\font minus \fontdimen4\font\relax}
\providecommand{\BIBforeignlanguage}[2]{{%
\expandafter\ifx\csname l@#1\endcsname\relax
\typeout{** WARNING: IEEEtran.bst: No hyphenation pattern has been}%
\typeout{** loaded for the language `#1'. Using the pattern for}%
\typeout{** the default language instead.}%
\else
\language=\csname l@#1\endcsname
\fi
#2}}
\providecommand{\BIBdecl}{\relax}
\BIBdecl

\bibitem{wagner2023dawn}
J.~Wagner, A.~Triantafyllopoulos, H.~Wierstorf, M.~Schmitt, F.~Burkhardt, F.~Eyben, and B.~W. Schuller, ``Dawn of the transformer era in speech emotion recognition: closing the valence gap,'' \emph{IEEE Transactions on Pattern Analysis and Machine Intelligence}, vol.~45, no.~9, pp. 10\,745--10\,759, 2023.

\bibitem{baevski2020wav2vec}
A.~Baevski, Y.~Zhou, A.~Mohamed, and M.~Auli, ``wav2vec 2.0: A framework for self-supervised learning of speech representations,'' \emph{Advances in neural information processing systems (NeurIPS)}, vol.~33, pp. 12\,449--12\,460, 2020.

\bibitem{harmon2017importance}
E.~Harmon-Jones, C.~Harmon-Jones, and E.~Summerell, ``On the importance of both dimensional and discrete models of emotion,'' \emph{Behavioral sciences}, vol.~7, no.~4, p.~66, 2017.

\bibitem{russell1979affective}
J.~A. Russell, ``Affective space is bipolar.'' \emph{Journal of personality and social psychology}, vol.~37, no.~3, p. 345, 1979.

\bibitem{lotfian2017buildingpodcast}
R.~Lotfian and C.~Busso, ``Building naturalistic emotionally balanced speech corpus by retrieving emotional speech from existing podcast recordings,'' \emph{IEEE Transactions on Affective Computing}, vol.~10, no.~4, pp. 471--483, 2017.

\bibitem{triantafyllopoulos2022probing}
A.~Triantafyllopoulos, J.~Wagner, H.~Wierstorf, M.~Schmitt, U.~Reichel, F.~Eyben, F.~Burkhardt, and B.~W. Schuller, ``Probing speech emotion eecognition transformers for linguistic knowledge,'' in \emph{Interspeech}, 2022.

\bibitem{micikevicius2017mixed}
P.~Micikevicius, S.~Narang, J.~Alben, G.~Diamos, E.~Elsen, D.~Garcia, B.~Ginsburg, M.~Houston, O.~Kuchaiev, G.~Venkatesh \emph{et~al.}, ``Mixed precision training,'' in \emph{International Conference on Learning Representations}, 2018.

\bibitem{hu2021lora}
E.~J. Hu, Y.~Shen, P.~Wallis, Z.~Allen-Zhu, Y.~Li, S.~Wang, L.~Wang, and W.~Chen, ``Lora: Low-rank adaptation of large language models,'' in \emph{International Conference on Learning Representations}, 2022.

\bibitem{vaessen2022fine}
N.~Vaessen and D.~A. Van~Leeuwen, ``Fine-tuning wav2vec2 for speaker recognition,'' in \emph{IEEE International Conference on Acoustics, Speech and Signal Processing (ICASSP)}, 2022, pp. 7967--7971.

\bibitem{yi2020applying}
C.~Yi, J.~Wang, N.~Cheng, S.~Zhou, and B.~Xu, ``Applying wav2vec2. 0 to speech recognition in various low-resource languages,'' \emph{arXiv preprint arXiv:2012.12121}, 2020.

\bibitem{chen2022large}
Z.~Chen, S.~Chen, Y.~Wu, Y.~Qian, C.~Wang, S.~Liu, Y.~Qian, and M.~Zeng, ``Large-scale self-supervised speech representation learning for automatic speaker verification,'' in \emph{IEEE International Conference on Acoustics, Speech and Signal Processing (ICASSP)}, 2022, pp. 6147--6151.

\bibitem{wang2021fine}
Y.~Wang, A.~Boumadane, and A.~Heba, ``A fine-tuned wav2vec 2.0/hubert benchmark for speech emotion recognition, speaker verification and spoken language understanding,'' \emph{arXiv preprint arXiv:2111.02735}, 2021.

\bibitem{pepino2021emotion}
L.~Pepino, P.~Riera, and L.~Ferrer, ``Emotion recognition from speech using wav2vec 2.0 embeddings,'' \emph{Interspeech}, 2021.

\bibitem{chen2023exploring}
L.-W. Chen and A.~Rudnicky, ``Exploring wav2vec 2.0 fine tuning for improved speech emotion recognition,'' in \emph{IEEE International Conference on Acoustics, Speech and Signal Processing (ICASSP)}, 2023, pp. 1--5.

\bibitem{eyben2015geneva}
F.~Eyben, K.~R. Scherer, B.~W. Schuller, J.~Sundberg, E.~Andr{\'e}, C.~Busso, L.~Y. Devillers, J.~Epps, P.~Laukka, S.~S. Narayanan \emph{et~al.}, ``The geneva minimalistic acoustic parameter set (gemaps) for voice research and affective computing,'' \emph{IEEE Transactions on Affective Computing}, vol.~7, no.~2, pp. 190--202, 2015.

\bibitem{pasad2021layer}
A.~Pasad, J.-C. Chou, and K.~Livescu, ``Layer-wise analysis of a self-supervised speech representation model,'' in \emph{IEEE Automatic Speech Recognition and Understanding Workshop (ASRU)}, 2021, pp. 914--921.

\bibitem{wu2022performance}
F.~Wu, K.~Kim, J.~Pan, K.~J. Han, K.~Q. Weinberger, and Y.~Artzi, ``Performance-efficiency trade-offs in unsupervised pre-training for speech recognition,'' in \emph{International Conference on Acoustics, Speech and Signal Processing (ICASSP)}, 2022, pp. 7667--7671.

\bibitem{lugo2024sustainable}
L.~Lugo and V.~Vielzeuf, ``Sustainable self-supervised learning for speech representations,'' \emph{arXiv preprint arXiv:2406.07696}, 2024.

\bibitem{feng2023peft}
T.~Feng and S.~Narayanan, ``Peft-ser: On the use of parameter efficient transfer learning approaches for speech emotion recognition using pre-trained speech models,'' in \emph{2023 11th International Conference on Affective Computing and Intelligent Interaction (ACII)}, 2023, pp. 1--8.

\bibitem{8855345}
E.~Naveen and P.~Kumar, ``Checkpointing in practice for memory-efficient training on the edge,'' in \emph{2019 IEEE 21st International Conference on High Performance Computing and Communications; IEEE 17th International Conference on Smart City; IEEE 5th International Conference on Data Science and Systems (HPCC/SmartCity/DSS)}, 2019, pp. 2759--2766.

\bibitem{sohoni2019low}
N.~S. Sohoni, C.~R. Aberger, M.~Leszczynski, J.~Zhang, and C.~R{\'e}, ``Low-memory neural network training: A technical report,'' \emph{arXiv preprint arXiv:1904.10631}, 2019.

\bibitem{loshchilov2017decoupled}
I.~Loshchilov and F.~Hutter, ``Decoupled weight decay regularization,'' in \emph{International Conference on Learning Representations}, 2019.

\bibitem{lawrence1989concordance}
I.~Lawrence and K.~Lin, ``A concordance correlation coefficient to evaluate reproducibility,'' \emph{Biometrics}, pp. 255--268, 1989.

\bibitem{parthasarathy2017jointly}
S.~Parthasarathy and C.~Busso, ``Jointly predicting arousal, valence and dominance with multi-task learning.'' in \emph{Interspeech}, 2017, pp. 1103--1107.

\bibitem{chen2022wavlm}
S.~Chen, C.~Wang, Z.~Chen, Y.~Wu, S.~Liu, Z.~Chen, J.~Li, N.~Kanda, T.~Yoshioka, X.~Xiao \emph{et~al.}, ``Wavlm: Large-scale self-supervised pre-training for full stack speech processing,'' \emph{IEEE Journal of Selected Topics in Signal Processing}, vol.~16, no.~6, pp. 1505--1518, 2022.

\bibitem{hsu2021hubert}
W.-N. Hsu, B.~Bolte, Y.-H.~H. Tsai, K.~Lakhotia, R.~Salakhutdinov, and A.~Mohamed, ``Hubert: Self-supervised speech representation learning by masked prediction of hidden units,'' \emph{IEEE/ACM transactions on audio, speech, and language processing}, vol.~29, pp. 3451--3460, 2021.

\end{thebibliography}

\end{document}